\documentclass[floatfix,aps,tightenlines,prd,nofootinbib,superscriptaddress,twocolumn,10pt]{revtex4-1}

\usepackage{color}
\usepackage{graphicx}
\usepackage{bm}
\usepackage{amsmath}
\usepackage{mathtools}

\newcommand{\Lk}{\ensuremath{\mathbf{L}^{\textrm{\scriptsize{-1}}}_{\textrm{k}}}}
\newcommand{\Rd}{\mathbf{R}_{\delta}}

\newcommand{\p}[1]{\mathbf{p}_{#1}}
\newcommand{\f}{\ensuremath{\mathbf{f}}}

\newcommand{\ti}[1]{\ensuremath{\mathbf{t}_{#1 } } }
\newcommand{\T}[1]{\ensuremath{\mathbf{T}_{#1 } } }

\newcommand{\Phio}{\ensuremath{\Phi_o}}
\newcommand{\Phimin}{\ensuremath{\Phi_{\textnormal{min}}}}
\newcommand{\Phimax}{\ensuremath{\Phi_{\textnormal{max}}}}

\begin{document}

\title{Neural mechanism to simulate a scale-invariant future}
\author{Karthik H.~Shankar, Inder Singh, and Marc W.~Howard}
\affiliation{Center for Memory and Brain, Initiative for the Physics and
		Mathematics of Neural Systems, Boston University}

\begin{abstract} 
Predicting future events, and their order, is important for efficient planning.
We propose a neural mechanism to non-destructively translate the current state
of memory into the future, so as to construct an ordered set of future
predictions.  This framework applies equally well to translations in time or
in one-dimensional position.  In a two-layer memory network that encodes the
Laplace transform of the external input in real time, translation can be
accomplished by modulating the weights between the layers. We propose that
within each cycle of hippocampal theta oscillations, the memory state is swept
through a range of translations to yield an ordered set of future predictions.
We operationalize several neurobiological findings into phenomenological
equations constraining translation.  Combined with constraints based on physical principles
requiring scale-invariance and coherence in translation across memory nodes, the proposition  results in Weber-Fechner spacing for the representation of both past
(memory) and future (prediction) timelines.  The resulting expressions are
consistent with findings from phase precession experiments in different
regions of the hippocampus  and reward systems in the ventral striatum.  The
model makes several experimental predictions that can be tested with existing
technology.
\end{abstract}

\maketitle{}

\section{Introduction}

The brain encodes externally observed stimuli in real time and represents
information about the current spatial location and temporal history of recent
events as activity distributed over neural networks.  Although we  are
physically localized in space and time, it is often useful for us to make
decisions based on non-local events, by anticipating events to occur at
distant future and remote locations. Clearly, a flexible access to  the
current state of spatio-temporal memory is crucial for the brain to
successfully anticipate events that might occur in the immediate next moment.
In order to anticipate events that might occur in the future after a given
time or at a given distance from the current location, the brain needs to
simulate how the current state of spatio-temporal memory representation will
have changed after waiting for a  given amount of time or after moving through
a given amount of distance. In this paper, we propose that the brain can
swiftly and non-destructively perform space/time-translation operations on the
memory state so as to anticipate events to occur at various future moments
and/or remote locations.

The rodent brain contains a rich and detailed representation of current spatial location and temporal history.  Some neurons--\emph{place cells}--in the hippocampus fire in circumscribed locations within an environment, referred to as their \emph{place fields}.  
Early work excluded confounds based on visual \cite{SaveEtal98}  or olfactory cues \cite{MullKubi87}, suggesting that the activity of place cells is a consequence of some form of path integration mechanism guided by the animal's velocity.   
Other neurons in the hippocampus---\emph{time cells}---fire during a
circumscribed period of time within a delay interval
\cite{PastEtal08,MacDEtal11,GillEtal11,KrauEtal13,Eich14}.  By analogy to
place cells, a set of time cells represents the animal's current temporal
position relative to past events.  Some researchers have long hypothesized a
deep connection between the hippocampal representations of place and time
\cite{EichCohe14,Eich00}.

\begin{figure}
 \includegraphics[width=0.98\columnwidth]{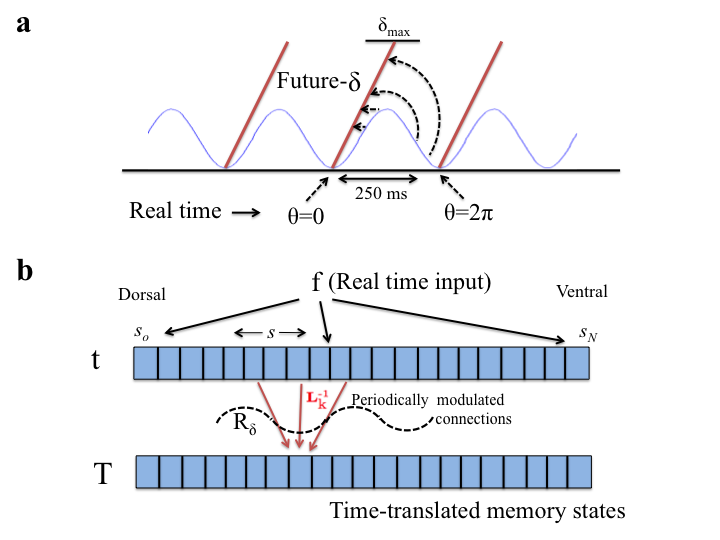} 
\caption{\textbf{a}. 
Theta oscillations of 4-8~Hz are observed in the voltage recorded from the hippocampus. \emph{Hypothesis}:  Within a theta cycle, a timeline of future translations  of magnitude $\delta$ is constructed.  \textbf{b}. Two layer network with theta-modulated connections. 
The $\ti{}$ layer receives external input $\f{}$ in real time and encodes its Laplace transform. The Laplace transform is inverted via a synaptic operator $\Lk$ to yield an estimate of the function $\f{}$ on the $\T{}$ layer nodes. By periodically manipulating the weights in $\Lk$, the memory state represented in $\T{}$ layer can be translated to represent its future states. 
 \label{fig:TILT}}
\end{figure}

Motivated by the spatial and temporal memory represented in the hippocampus,
we hypothesize that the translation operation required to anticipate the
events at a distant future engages this part of the brain
\cite{HassEtal07,SchaEtal07}. We hypothesize
that \emph{theta oscillations}, a well-characterized rhythm of 4-8 Hz in the
local field potential observed in the hippocampus may be responsible for the
translation operation.  In particular, we hypothesize that sequential
translations of different magnitudes take place at different phases  within a
cycle of theta oscillation, such that a timeline of anticipated future events
(or equivalently a spaceline of anticipated events at distant locations) is
swept out in a single cycle (fig.~\ref{fig:TILT}a).

Theta oscillations are prominently observed during periods of navigation
\cite{Vand69}. Critically, there is a systematic relationship between the
animal's position within a neuron's place field and the phase of the theta
oscillation at which that neuron fires \cite{OKeeRecc93}, known as
\emph{phase precession}.
This suggests that the phase of firing of the place cells conveys
information about the anticipated future location of the animal. This provides a
strong motivation for our hypothesis that  the phase of theta oscillation would be
linked to the translation operation.

\subsection{Overview}

This paper develops a computational mechanism for the translation operation of
a spatial/temporal memory representation constructed from a two-layer neural
network model \cite{ShanHowa12}, and links it to theta oscillations by
imposing certain constraints based on some neurophysiological observations and
some physical principles we expect the brain to satisfy. Since the focus here
is to understand the computational mechanism of a higher level cognitive
phenomena, the imposed constraints and the resulting derivation should
be viewed at a phenomenological level, and not as emerging from biophysically
detailed neural interactions.

Computationally, we assume that the memory representation is constructed by a
two-layer network (fig.~\ref{fig:TILT}b) where the first layer encodes the
Laplace transform of externally observed stimuli in real time, and the second
layer approximately inverts the Laplace transform to represent a fuzzy
estimate of the actual stimulus history \cite{ShanHowa12}.   With access to
instantaneous velocity of motion, this two layer network representing temporal
memory can be straightforwardly generalized to represent one-dimensional
spatial memory \cite{HowaEtal14}. Hence in the context of this two layer
network, time-translation of the temporal memory representation can be
considered mathematically equivalent to space-translation of the spatial
memory representation. 

 Based on a simple, yet powerful, mathematical observation that translation
 operation can be performed in the Laplace domain as an instantaneous
 point-wise product, we propose that the translation operation is achieved by
 modulating the connection weights between the two layers within each theta
 cycle (fig.~\ref{fig:TILT}b). The translated representations can then be used
 to predict events at distant future and remote locations.  In constructing
 the translation operation, we impose two physical principles we expect the
 brain to satisfy. The first principle is \emph{scale-invariance}, the
 requirement that all scales (temporal or spatial) represented in the memory
 are treated equally in implementing the translation. The second principle is
 \emph{coherence}, the requirement that at any moment all nodes forming the
 memory representation are in sync, translated by the same amount.

Further, to implement the computational mechanism of translation as a neural
mechanism, we impose certain phenomenological constraints based on
neurophysiological observations. First, there exists a dorsoventral axis in
the hippocampus of a rat's brain, and the size of place fields increase
systematically from the dorsal to the ventral end \cite{JungEtal94,KjelEtal08}.
In light of this observation, we hypothesize that the nodes representing
different temporal and spatial scales of memory are ordered along the dorsoventral
axis. Second, the phase of theta oscillation is not uniform along the
dorsoventral axis; phase advances from the dorsal to the ventral end like
a traveling wave \cite{LubeSiap09,PateEtal12} with a phase difference of about
$\pi$ from one end to the other. Third, the synaptic weights change
as a function of phase of the theta oscillation throughout the hippocampus
\cite{WyblEtal00,SchaEtal08}. In light of this observation, we hypothesize
that the change in the connection strengths between the two layers required to
implement the translation operation depend only on the local phase of the theta oscillation at any  node (neuron).

In section~\ref{sec:math}, we impose the above mentioned physical principles
and phenomenological constraints to derive quantitative relationships for the
distribution of scales of the nodes representing the memory and the
theta-phase dependence of the translation operation. This yields specific
forms of phase-precession in the nodes representing the  memory as well as the
nodes representing future prediction.  Section~\ref{sec:nbio} compares these
forms to neurophysiological phase precession observed in the hippocampus and
ventral striatum.  Section~\ref{sec:nbio} also makes explicit
neurophysiological predictions that could verify our hypothesis that theta
oscillations implement the translation operation to construct a timeline of
future predictions.

\section{Mathematical model}
\label{sec:math}

In this section we start with a basic overview of the two layer memory model
and summarize the relevant details from previous work
\cite{ShanHowa12,ShanHowa13,HowaEtal14} to serve as a background.  Following
that,  we derive the equations that allow the memory nodes to be coherently
time-translated to various future moments in synchrony with the theta
oscillations. Finally we derive the predictions generated for various future
moments from the time-translated memory states.

\subsection{Theoretical background}

\label{ssec:mathback}

The memory model is implemented as  a two-layer feedforward network
(fig.~\ref{fig:TILT}b) where the $\ti{}$ layer holds the Laplace transform of
the recent past and the $\T{}$ layer reconstructs  a temporally fuzzy estimate
of past events \cite{ShanHowa12,ShanHowa13}.  Let the stimulus at any time
$\tau$ be denoted as  $\f{}(\tau)$. The nodes in the $\ti{}$ layer  are leaky
integrators parametrized by their decay rate $s$, and are all independently
activated by the stimulus.  The nodes are assumed to be arranged w.r.t.~their
$s$ values.  The nodes in the $\T{}$ layer are in one to one correspondence
with the nodes in the $\ti{}$ layer and hence can also be parametrized by the
same $s$. The feedforward connections from the $\ti{}$ layer into the $\T{}$
layer are prescribed to satisfy certain mathematical properties which are
described below. The activity of the two layers is given by 
\begin{eqnarray}
  \frac{d}{d \tau} \ti{}(\tau,s) &=& -s \ti{}(\tau,s) + \f{}(\tau) \label{eq:teq} \\ 
  \T{}(\tau,s) &=&  [\Lk ]  \, \ti{}(\tau,s) \label{eq:Teq} 
\end{eqnarray}   
By integrating  eq.~\ref{eq:teq}, note that  the $\ti{}$ layer encodes the
Laplace transform of the entire past of the stimulus function leading up to
the present. The $s$ values distributed over the $\ti{}$ layer represent  
the (real) Laplace domain variable.  The fixed connections between the $\ti{}$
layer and $\T{}$ layer denoted by the operator $\Lk$ (in eq.~\ref{eq:Teq}), is
constructed to reflect an approximation to inverse Laplace transform.  In
effect, the Laplace transformed  stimulus history which is distributed over
the $\ti{}$ layer nodes is inverted by $\Lk$ such that a fuzzy (or coarse
grained) estimate of the actual stimulus value from various
past moments is represented along the different $\T{}$ layer nodes. 

More precisely, treating the $s$ values nodes as  continuous, the $\Lk$ operator can be succinctly expressed as 
\begin{equation}
\T{}(\tau,s) = \frac{(-1)^k}{k!} s^{k+1} \ti{}^{(k)} (\tau,s) \equiv [\Lk]  \, \ti{}(\tau,s)
\label{eq:Lk}
\end{equation}
Here $\ti{}^{(k)}(\tau,s)$ corresponds to the $k$-th derivative of
$\ti{}(\tau,s)$ w.r.t.~$s$.  It can be proven that $\Lk$ operator executes an
approximation to the inverse Laplace transformation and the approximation
grows more and more accurate for larger and larger values of $k$
\cite{Post30}.  Further details of  $\Lk$ depends on the $s$ values chosen for
the nodes \cite{ShanHowa13}, but these details are not relevant for this
paper as the $s$ values of neighboring nodes are assumed to be close enough
that the analytic expression for $\Lk$ given by eq.~\ref{eq:Lk} would be
accurate.

To emphasize the properties of this memory representation, consider the
stimulus $\f{}(\tau)$ to be a Dirac delta function at $\tau=0$. From
eq.~\ref{eq:teq}~and~\ref{eq:Lk}, the  $\T{}$ layer activity following the
stimulus presentation ($\tau >0$) turns out to be      
\begin{equation}
\T{} (\tau, s) = \frac{s}{k!}  \left[s\tau \right]^k
e^{-\left[s\tau \right]}
\label{eq:old_big_T}
\end{equation}
Note that nodes with different $s$ values in the $\T{}$ layer peak in activity after different delays following the stimulus; hence the $\T{}$ layer nodes behave like {time cells}. In particular, a node with a given $s$ peaks in activity at a time $\tau=k/s$ following the stimulus. Moreover, viewing the activity of any node as a distribution around its appropriate peak-time ($k/s$), we see that the shape of this distribution is  exactly the same for all nodes to the extent $\tau$ is rescaled to align the peaks of all the nodes. In other words, the activity of different nodes of the $\T{}$ layer represent a fuzzy estimate of the past information from different timescales and the fuzziness associated with them is directly proportional to the timescale they represent, while maintaining the exact same shape of fuzziness. For this reason, the $\T{}$ layer represents the past information in a {scale-invariant} fashion.

This two-layer memory architecture is also amenable to represent
one-dimensional spatial memory analogous to the representation of temporal
memory in the $\T{}$ layer \cite{HowaEtal14}. If the stimulus $\f{}$ is
interpreted as a landmark encountered at a particular location in a
one-dimensional spatial arena, then the $\ti{}$ layer nodes can be made to
represent the Laplace transform of the landmark treated as a spatial function
with respect to the current location. By modifying eq.~\ref{eq:teq} to  
\begin{equation}
 \frac{d}{d \tau} \ti{}(\tau,s) = v \left[ -s \ti{}(\tau,s) + \f{}(\tau) \right]  , 
 \label{eq:timespace}
\end{equation}
where $v$ is the velocity of motion, the temporal dependence of the $\ti{}$ layer activity can be converted to spatial dependence.\footnote{ Theoretically, the velocity here could  be an animal's running velocity in the
lab maze or a mentally simulated human motion while playing video games.} By employing the $\Lk$ operator on this modified $\ti{}$ layer activity (eq.~\ref{eq:timespace}), it is straightforward to construct a layer of nodes (analogous to $\T{}$) that exhibit peak activity at different distances from the landmark. Thus the two-layer memory architecture can be trivially extended to yield {place-cells} in one dimension. 

In what follows, rather than referring to translation operations separately on spatial and temporal memory,  we shall simply consider time-translations with an implicit understanding that all the results derived
can be trivially extended to 1-d spatial memory representations.

\subsection{Time-translating the Memory state}

\label{sec:newstuff}

The two-layer architecture naturally lends
itself for {time-translations} of the memory state in the $\T{}$ layer,
which we shall later exploit to construct a timeline of future predictions.
The basic idea is that if the current state of memory represented in the
$\T{}$ layer is used to anticipate the present (via some prediction
mechanism), then a time-translated state of $\T{}$ layer can be used to predict
events that will occur at a distant future via the same prediction
mechanism. Time-translation means to mimic the $\T{}$ layer
activity at a distant future based on its current state. Ideally translation
should be {non-destructive}, not overwriting the current activity in
the $\ti{}$ layer.

Let $\delta$ be the amount by which we intend to time-translate the state of
$\T{}$ layer. So, at any time $\tau$, the aim is to access
$\T{}(\tau+\delta,s)$ while still preserving the current $\ti{}$ layer
activity, $\ti{}(\tau,s)$.  This is can be easily achieved because the $\ti{}$
layer represents the stimulus history in the Laplace domain. Noting that the
Laplace transform of a $\delta$-translated function is simply the product of
$e^{-s \delta}$ and the Laplace transform of the un-translated function, we see
that   
\begin{eqnarray}
\ti{}(\tau+\delta,s) &=& e^{-s\delta} \ti{}(\tau,s) \label{eq:translate}
\end{eqnarray}   
Now noting that $\T{}(\tau+\delta,s)$ can be obtained by employing the $\Lk$
operator on $\ti{}(\tau+\delta,s)$ analogous to eq.~\ref{eq:Lk}, we obtain the
$\delta$-translated $\T{}$ activity as 
\begin{eqnarray}
\T{\delta}(\tau,s) \equiv \T{}(\tau+\delta,s) &=& [ \Lk]  \,  \ti{}(\tau+\delta ,s)  \nonumber \\
  &=& \left[\Lk \cdot \Rd \right]  \, \ti{}(\tau,s) 
  \label{eq:Rd}  
\end{eqnarray}  
where $\Rd$ is just a diagonal operator whose rows and columns are indexed by
$s$ and the diagonal entries are $e^{-s \delta}$. The $\delta$-translated
activity of the $\T{}$ layer is now subscripted by $\delta$ as $\T{\delta}$ so
as to distinguish it from the un-translated $\T{}$ layer activity given by
eq.~\ref{eq:Lk} without a subscript. In this notation the un-translated state
$\T{}(\tau,s)$ from eq.~\ref{eq:Lk} can be expressed as $\T{0}(\tau,s)$. The
time-translated $\T{}$ activity can be obtained from the current $\ti{}$ layer
activity if the connection weights between the two layers given by $\Lk$ is
modulated by $\Rd$. This computational mechanism of time-translation can be implemented as a neural mechanism in the brain, by imposing certain phenomenological constraints and physical principles. 

\begin{figure}
\includegraphics[width=0.4\textwidth]{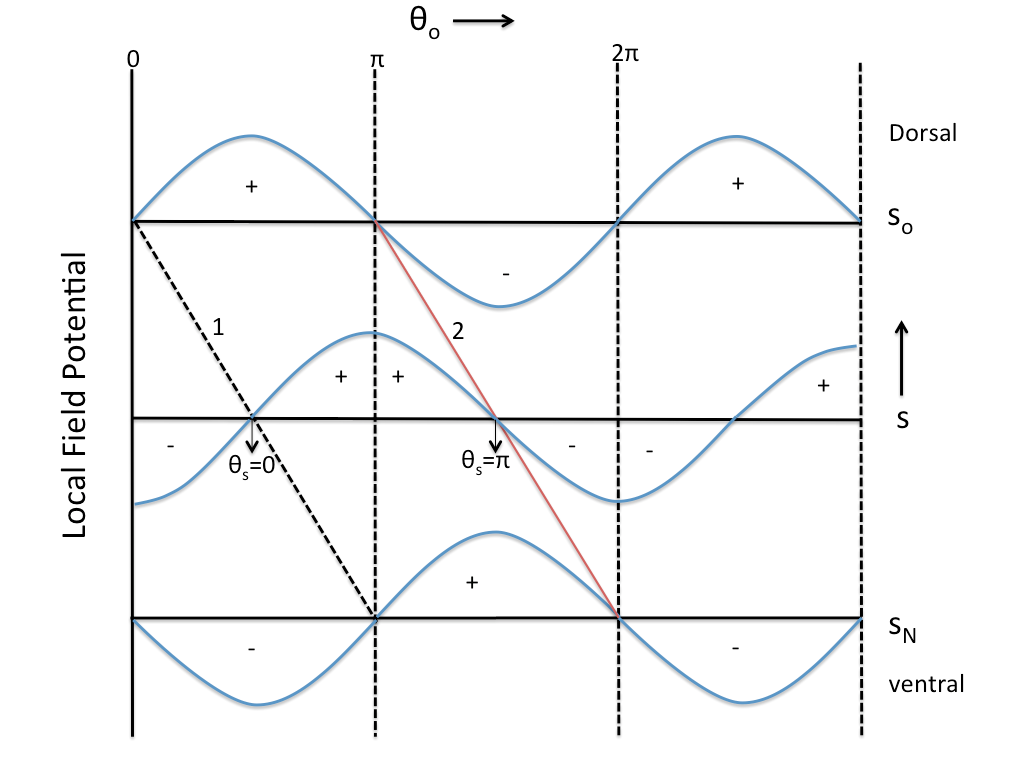}
\caption{Traveling theta wave along the $s$ axis. 
The x-axis is real time.  Each point along the dorsoventral axis corresponds
		to a different value of $s_n$.  The curvy blue lines show the theta
		oscillation for several different values of $s$.  Lines~1~and~2
		connect the  positions where the local phases $\theta_s$ are 0 and
		$\pi$ respectively.
		\label{fig:theta}}
\end{figure}

{\bf \emph{Observation 1:}} 
Anatomically, along the dorsoventral axis of the hippocampus, the width of
place fields systematically increases from the dorsal end to the
ventral end \cite{JungEtal94,KjelEtal08}.  Fig.~\ref{fig:theta} schematically
illustrates this observation by identifying the $s$-axis of the two-layer
memory architecture with the dorso-ventral axis of the hippocampus,
such that the scales represented by the nodes are monotonically arranged.  Let
there be $N+1$ nodes with monotonically decreasing $s$ values given by $s_o$,
$s_1$, \ldots $s_N$.  

{\bf \emph{Observation 2:}} 
The phase of the theta oscillations along the axis is non-uniform,
representing a traveling wave from the dorsal to ventral part of the hippocampus
with a net phase shift of $\pi$ \cite{LubeSiap09,PateEtal12}. 
The oscillations  in fig.~\ref{fig:theta} symbolize the local
field potentials at different locations of the $s$-axis.  The local phase of
the oscillation at any position on the $s$-axis is denoted by $\theta_s$,
which ranges from $-\pi$ to $+\pi$ by convention. However, as a reference we
denote the phase at the top (dorsal) end as $\theta_o$ ranging from $0$ to $2
\pi$, with the understanding that the range $(\pi, 2 \pi)$ is mapped on to
$(-\pi, 0)$. The $x$-axis in fig.~\ref{fig:theta} is time within a theta
oscillation labeled by the phase $\theta_o$.

In this convention, the value of $\theta_s$ discontinuously jumps from $+\pi$
to $-\pi$ as we move from one cycle of oscillation to the next. In
fig.~\ref{fig:theta}, the diagonal (solid-red) line labeled `2' denotes all
the points where this discontinuous jump happens. The diagonal (dashed) line
labeled  `1' denotes all the points where $\theta_s=0$. It is straightforward
to infer the relationship between the phase at any two values of $s$. Taking
the nodes to be uniformly spaced anatomically, the local phase $\theta_s$ of the $n$-th
node is related to $\theta_o$ (for $0<\theta_s<\pi$) by\footnote{ Since the $s$
values of the nodes are monotonically arranged, we can interchangeably use $s$
or $n$ as subscritpts to $\theta$. } 
\begin{equation}
\theta_s/\pi= \theta_o/\pi - n/N .  
\label{eq:thetasnophi}
\end{equation}

{\bf \emph{Observation 3:}} Synaptic weights in the hippocampus are modulated
periodically in synchrony with the phase of  theta
oscillation \cite{WyblEtal00,SchaEtal08}. Based on this observation, we impose
the constraint that the connection strengths between the $\ti{}$ and $\T{}$
layers at a particular value of $s$ depend only on the local phase of the
theta oscillations. Thus the diagonal entries in the $\Rd$ operator should
only depend on $\theta_s$. We take  these entries to be of the form $\exp{
		(-\Phi_s (\theta_s)) } $, where $\Phi_s$ is any continuous function of
		$\theta_s  \in (-\pi, +\pi)$. Heuristically, at  any moment
		within a theta cycle, a $\T{}$ node with a given $s$ value will be
		roughly translated by an amount $\delta = \Phi_s(\theta_s)/s$. 

{\bf \emph{Principle 1:}} \emph{Preserve Scale-Invariance}

Scale-invariance is an extremely adaptive property for a memory to have; in
many cases biological memories seem to exhibit scale-invariance
\cite{BalsGall09}.
As the untranslated $\T{}$ layer activity already exhibits scale-invariance, we impose the constraint
that the time-translated states of $\T{}$ should also exhibit scale-invariance.  This consideration requires the behavior of every node to follow the same pattern with respect to their local theta phase. This amounts
to choosing the functions $\Phi_s$ to be the same for all $s$, which we shall refer to as $\Phi$.   

{\bf \emph{Principle 2:}} \emph{Coherence in translation}

Since the time-translated memory state is going to be used to make predictions
for various moments in the distant future, it would be preferable if all the nodes
are time-translated by the same amount at any moment within a theta
cycle. If not, different nodes would contribute to predictions for different
future moments leading to noise in the prediction.  However, such a
requirement of global coherence cannot be imposed consistently along with the
principle~1 of preserving scale-invariance.\footnote{This is easily seen by
noting that each node will have a maximum translation inversely
proportional to its $s$-value to satisfy principle~1.} But in the light of
prior work \cite{HassEtal02,Hass12} which suggest that retrieval of
memory or prediction happens only in one half of the theta cycle,\footnote{
This hypothesis follows from the observation that while both synaptic
transmission and synaptic plasticity are modulated by theta phase,
they are out of phase with one another.  That is, while certain synapses are
		learning, they are not communicating information and vice versa.  This
		led to the hypothesis that the phases where plasticity is
optimal are specialized for encoding whereas the phases where transmission is
optimal are specialized for retrieval.  }
we 
impose the requirement of coherence only to those nodes that are all in the
positive half of the cycle at any moment. That is, $\delta= \Phi(\theta_s)/s$
is a constant along any vertical line in the region bounded between the
diagonal lines~1~and~2 shown in fig.~\ref{fig:theta}.  Hence for all nodes
with $0<\theta_s< \pi$, we require 
\begin{equation}
\Delta \left(\Phi\left(\theta_s\right) / s \right) = \Delta \left(\Phi\left(\theta_o - \pi n/N\right) / s_n \right)  = 0.
\label{eq:deltathetas}
\end{equation}

For coherence as expressed in eq.~\ref{eq:deltathetas} to hold at all values
of $\theta_o$ between 0 and $2 \pi$, $\Phi(\theta_s)$ must be an exponential
function so that $\theta_o$ can be
functionally decoupled from $n$; consequently $s_n$ should also have an
exponential dependence on $n$. So the general solution to
eq.~\ref{eq:deltathetas} when $0<\theta_s<\pi$ can be written as
\begin{eqnarray}
\Phi(\theta_s)&=& \Phio \exp{[b \theta_s]}   \label{eq:thetas}  \\
   s_n &=& s_o (1+c)^{-n} \label{eq:thetasn}
\end{eqnarray} 
where $c$ is a positive number. In this paper, we shall take $c \ll 1$, so
that the analytic approximation for the $\Lk$ operator given in terms of the
$k$-th derivative along the $s$ axis in  eq.~\ref{eq:Lk} is valid. 

Thus the requirement of coherence in time-translation implies that the $s$
values of the nodes---the timescales represented by the nodes---are spaced out
exponentially, which can be referred to as a Weber-Fechner scale, a commonly
used terminology in cognitive science. Remarkably, this result strongly
resonates with a requirement of the exact same scaling when the predictive
information contained in the memory system is maximized in response to 
long-range correlated signals \cite{ShanHowa13}.  This feature allows this
memory system to represent scale-invariantly coarse grained past information
from timescales exponentially related to the number of nodes. 

The maximum value attained by the function $\Phi( \theta_s)$ is at $\theta_s=\pi$, and the maximum value is $\Phimax = \Phio \exp{[b \pi] }$, such that $\Phimax/\Phio = s_o/s_N$ and   $b=(1/\pi) \log{(\Phimax/\Phio)}$.
To ensure continuity around $\theta_s=0$, we take the eq.~\ref{eq:thetas} to
hold true even for $\theta_s \in (-\pi,0)$. However, since notationally
$\theta_s$ makes a  jump from $+\pi$ to $-\pi$, $\Phi(\theta_s)$ would exhibit
a discontinuity at the diagonal line~2 in fig.~\ref{fig:theta} from $\Phimax$
(corresponding to $\theta_s=\pi $) to $\Phimin=\Phio^2/\Phimax$
(corresponding to $\theta_s=-\pi$).

Given these considerations, at any instant within a theta cycle, referenced by
the phase $\theta_o$, the amount $\delta$ by which the memory state is
time-translated can be derived from  eq.~\ref{eq:thetasnophi} and
\ref{eq:thetas} as
\begin{equation}
\delta(\theta_o) = (\Phio/s_o)  \exp{[b \theta_o]}  .
\label{eq:deltatheta}
\end{equation}
Analogous to having the past represented on a Weber-Fechner scale, the
translation distance $\delta$ into the future also falls on a Weber-Fechner
scale as the theta phase is swept from 0 to $2 \pi$. In other words, the
amount of time spent within a theta cycle for larger translations is
exponentially smaller.

To emphasize the properties of the time-translated $\T{}$ state, consider the
stimulus to be a Dirac delta function at $\tau=0$. From eq.~\ref{eq:Rd}, we
can express the $\T{}$ layer activity analogous to eq.~\ref{eq:old_big_T}.
\begin{equation}
\T{\delta} (\tau, s) \simeq \frac{s}{k!}  \left[s\tau+\Phi\left(\theta_s\right)\right]^k
e^{-\left[s\tau+ \Phi\left(\theta_s\right)\right]}
\label{eq:bigT}
\end{equation}
Notice that eqs.~\ref{eq:thetasnophi} and \ref{eq:deltatheta} specify a unique
relationship between  $\delta$ and $\theta_s$ for any given $s$.  The 
r.h.s.~above is expressed in terms of $\theta_s$ rather than $\delta$ so as to shed
light on the phenomenon of {phase precession}. 

Since $\T{\delta}(\tau,s)$ depends on both $\tau$ and $\theta_s$ only via the
sum $[ s\tau+ \Phi\left(\theta_s\right) ] $, a given node will show identical
activity for various combinations of $\tau$ and $\theta_s$.\footnote{While
representing timescales much larger than the period of a theta cycle, $\tau$
can essentially be treated as a constant within a single cycle. In other
words,  $\theta_s$ and $\tau$ in eq.~\ref{eq:Rd} can be treated as
independent, although in reality the phases evolve in real time.} 
For instance, a node would achieve its peak activity when $\tau$ is
significantly smaller than its timescale $(k/s)$ only when $\Phi(\theta_s)$ is
large---meaning $\theta_s \simeq +\pi$. And as $\tau$ increases towards the
timescale of the node, the peak activity gradually shifts to earlier phases
all the way to  $\theta_s \simeq -\pi$.  An important consequence of imposing
principle~1 is that the relationship between $\theta_s$ and $\tau$ on any
iso-activity contour is scale-invariant. That is, every node behaves similarly when 
$\tau$ is rescaled by the timescale of the node.  We shall
further pursue the analogy of this phenomenon of phase precession with
neurophysiological findings  in the next section (fig.~\ref{fig:ph-pr}).

\subsection{Timeline of Future Prediction}
				
At any moment, $\T{\delta}$ (eq.~\ref{eq:bigT}) can be used to predict the
stimuli expected at a future moment. Consequently, as $\delta$
is swept through within a theta cycle, a timeline of future predictions can be
simulated in an orderly fashion, such that predictions for closer events occur
at earlier phases (smaller $\theta_o$) and predictions of distant events occur
at later phases. In order to predict from a time-translated state
$\T{\delta}$, we need a prediction mechanism.
For our purposes, we consider here a very simple form of learning and
prediction, \emph{Hebbian association}.  In this view, an event is learned (or
an association formed in long term memory) by increasing the connection
strengths between the neurons representing the currently-experienced
stimulus and the neurons representing the recent past events ($\T{0}$).
Because the $\T{}$ layer activity contains temporal information about the preceding stimuli, simple associations between $\T{0}$ and the current stimulus
are sufficient to encode and express well-timed predictions \cite{ShanHowa12}.
In particular, the term Hebbian implies that the change in each connection
strength is proportional to the product of pre-synaptic activity---in this
case the activity of the corresponding node in the $\T{}$ layer---and
post-synaptic activity corresponding to the current stimulus.
Given that the associations are learned in this way, we define the prediction of a
particular stimulus to be the scalar product of its association strengths with 
the current state of $\T{}$. In this way, the scalar product of association 
strengths and a translated state $\T{\delta}$ can be understood as the
future prediction of that stimulus.

Consider the thought experiment where a
conditioned stimulus \textsc{cs} is consistently followed by another
stimulus, \textsc{a} or \textsc{b}, after a time $\tau_o$. Later when
\textsc{cs} is repeated (at a time $\tau=0$), the subsequent activity in the
$\T{}$ nodes can be used to generate predictions for the future occurrence of
\textsc{a} or \textsc{b}.  The connections to the node corresponding to
\textsc{a} will be incremented by the state of $\T{0}$ when \textsc{a} is
presented; the connections to the node corresponding to
\textsc{b} will be incremented by the state of $\T{0}$ when \textsc{b} is
presented.  In the context of Hebbian learning, the prediction for the
stimulus at a future time as a function of $\tau$ and $\tau_o$ is obtained as
the sum of $\T{\delta}$ activity of each node multiplied by the learned
association strength ($\T{0}$):  
\begin{equation}
\p{\delta}(\tau,\tau_o) = \sum_{n=\ell}^{N} \T{\delta}\left(\tau,s_n\right)
	\ \T{0}\left(\tau_o,s_n\right)/  s_{n}^{w}.
	\label{eq:pthetatau}
\end{equation}
The factor $s_n^w$ (for any $w$)  allows for differential association
strengths for the different $s$ nodes, while still preserving the scale
invariance property.  Since $\delta$ and $\theta_o$ are monotonically related
(eq.~\ref{eq:deltatheta}), the prediction $\p{\delta}$ for various future moments
happens at various phases of a theta cycle. 

Recall that all the nodes in the $\T{}$ layer are coherently time-translated
only in the positive half of the theta cycle. Hence for computing future
predictions based on a time-translated state $\T{\delta}$, only coherent nodes
should contribute. In fig.~\ref{fig:theta}, the region to the right of 
diagonal line~2 does not contribute to the prediction. The lower
limit $\ell$ in the summation over the nodes given in eq.~\ref{eq:pthetatau}
is the position of the diagonal line~2 in fig.~\ref{fig:theta} marking the
position of discontinuity where $\theta_s$ jumps from $+\pi$ to $-\pi$.    

In the limit when $c \rightarrow 0$, the $s$ values of neighboring nodes are
very close and the summation can be approximated by an integral.  Defining
$x = s \tau_o$ and $y= \tau/\tau_o$ and $v= \delta/\tau_o$, the
 above summation can be rewritten as 
 \begin{equation}
\p{\delta}(\tau,\tau_o) \simeq \frac{\tau_o^{w-2}}{k!^2}  \int_{x_{min}}^{x_u} x^{2k+1-w} (y +v)^k e^{-x(1+y+v)} \, dx
\end{equation}
Here $x_{min} = s_N \tau_o$, and $x_u = s_o \tau_o$ for $0< \theta_o< \pi$ and $x_u= \Phi_{max} \tau_o/\delta$ for $\pi< \theta_o <2 \pi$. The integral can be evaluated in terms of lower incomplete gamma functions to be
 \begin{eqnarray}
&\p{\delta}&(\tau,\tau_o) \simeq \frac{\tau_o^{w-2}}{k!^2}
		\frac{\left[ (\tau+\delta )/\tau_o \right]^k}{[1+ (\tau+\delta)/\tau_o]^C}  \,
		\times \nonumber \\
  & & \left(  \Gamma \left[C, (\tau_o+\tau +\delta) U \right] - \Gamma \left[C, (\tau_o+\tau +\delta) s_N \right] \right),
 \label{eq:p}   
\end{eqnarray} 
where $C=2k+2-w$ and $\Gamma[.,.]$ is the lower incomplete gamma function.
For $\theta_o<\pi$ (i.e., when $\delta < \Phimax/s_o$), $U=s_o$ and for
$\theta_o>\pi$  (i.e., when $\delta > \Phimax/s_o$), $U=\Phimax/\delta$.

\begin{figure}
\includegraphics[width=0.5\textwidth]{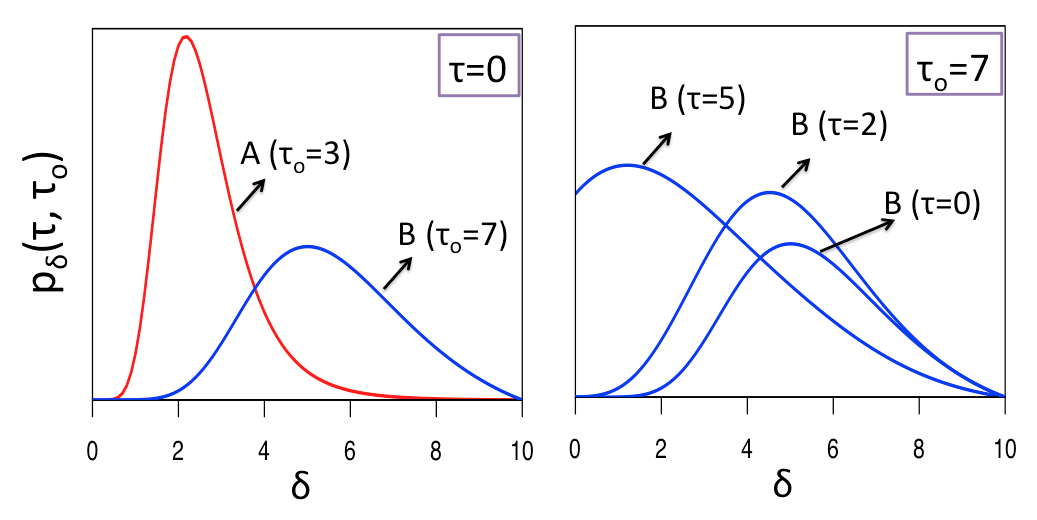}
\caption{Future timeline. Eq.~\ref{eq:p} is plotted as a function of $\delta$.  During training, the \textsc{cs} was presented at $\tau_o=3$ before \textsc{a} and $\tau_o=7$ before \textsc{b}. Left: Immediately after presentation of the \textsc{cs}, the predictions for \textsc{a} and \textsc{b} are ordered on the $\delta$
axis.   Note that the prediction for \textsc{b} approximates a rescaled
		version of that for  \textsc{a}.  Right: The prediction for \textsc{b}
is shown for varying times after presentation of \textsc{cs}.  With the
		passage of time, the prediction of \textsc{b} becomes stronger and
		more imminent.  In this figure,
			 $\Phimax=10$, $\Phio=1$, $k=10$, $s_o=10$, $s_N=1$, and $w=1$.
					 \label{fig:futurepred}
}
\end{figure}

Figure~\ref{fig:futurepred} provides a graphical representation of some key
properties of eq.~\ref{eq:p}.  The figure assumes that the \textsc{cs} is
followed by \textsc{a} after $\tau_o=3$ and followed by \textsc{b} after
$\tau_o=7$.  The left panel shows the predictions for both \textsc{a} and
\textsc{b}  as a function of $\delta$ immediately after presentation of
\textsc{cs}.  The prediction for \textsc{a} appears at smaller
$\delta$ and with a higher peak than the prediction for
\textsc{b}.  The value of $w$ affects the relative sizes of the peaks.  The
right panel shows how the prediction for \textsc{b} changes with the passage
of time after presentation of the \textsc{cs}.  As $\tau$ increases from zero
and the \textsc{cs} recedes into the past, the prediction of \textsc{b} peaks
at smaller values of $\delta$, corresponding to more imminent future times.
In particular when $\tau_o$ is much smaller than the largest (and larger than
the smallest) timescale represented by the nodes, then the
shape of $\p{\delta}$ remains the same when $\delta$ and $\tau$ are rescaled by
$\tau_o$. Under these conditions, the timeline of future predictions generated
by $\p{\delta}$ is scale-invariant.

Since $\delta$ is in one-to-one relationship with $\theta_o$, as a predicted
stimulus becomes more imminent, the activity corresponding to that predicted
stimulus should peak at earlier and earlier phases. Hence a timeline of future
predictions can be constructed from  $\p{\delta}$ as the phase $\theta_o$ is
swept from $0$ to $2 \pi$. Moreover the cells representing $\p{\delta}$ should
show phase precession with respect to $\theta_o$. Unlike cells representing
$\T{\delta}$, which depend directly on their local theta phase, $\theta_s$,
the phase precession of  cells representing $\p{\delta}$ should depend
on the reference phase $\theta_o$ at the dorsal end of the $s$-axis.
We shall further connect this neurophysiology in the next section
(fig.~\ref{fig:MeerRedi}).

\section{Comparisons with Neurophysiology}
\label{sec:nbio}

The mathematical development focused on two entities $\T{\delta}$ and $\p{\delta}$ that change their value based on the theta phase (eqs.~\ref{eq:bigT}~and~\ref{eq:p}).  In order to compare these to neurophysiology, we need to have some hypothesis linking them to the activity of neurons from specific brain regions. We emphasize that although the development in the preceding section was done
with respect to time, all of the results generalize to one-dimensional position as well (eq.~\ref{eq:timespace}, \cite{HowaEtal14}). The overwhelming majority of evidence for phase precession comes from studies of place cells (but see \cite{PastEtal08}).  Here we compare the properties of $\T{\delta}$ to phase precession in hippocampal neurons and the properties of $\p{\delta}$ to a study showing phase precession in
ventral striatum \cite{MeerEtal11}.\footnote{This is not meant to preclude the possibility that $\p{\delta}$ could  be computed at other parts of the brain as well.}

Due to various analytic approximations, the activity of nodes in the $\T{}$
layer as well as the activity of the nodes representing future prediction
(eqs.~\ref{eq:bigT}~and~\ref{eq:p}) are expressed as smooth functions of time
and theta phase. However, neurophysiologically, discrete spikes (action
				potentials) are observed.  In order to facilitate comparison
of the model to neurophysiology, we adopt a simple stochastic spike-generating
method. In this simplistic approach, the activity of the nodes given by
eqs.~\ref{eq:bigT}~and~\ref{eq:p} are taken to be proportional to the
instantaneous probability for generating a spike. The probability of
generating a spike at any instant is taken to be the instantaneous activity
divided by the maximum activity achieved by the node if the activity is
greater than 60\% of the maximum activity.  In addition, we add spontaneous
stochastic spikes at any moment with a probability of 0.05. For all of the
figures in this section, the parameters of the model are set as $k=10$,
		$\Phimax=10$, $w=2$, $\Phi_o = 1$, $s_N=1$, $s_o=10$.

This relatively coarse level of realism in spike generation from the analytic
expressions is probably appropriate to the resolution of the experimental
data.   There are some experimental challenges associated with exactly
evaluating the model.  First, theta phase has to be estimated from a noisy
signal.  Second, phase precession results are typically shown as averaged
across many trials. It is not necessarily the case that the average is
representative of an individual trial (although this is the case at least for
phase-precessing cells in medial entorhinal cortex \cite{ReifEtal12}).
Finally, the overwhelming majority of phase precession experiments utilize
extracellular methods, which cannot perfectly identify spikes from individual
neurons.

\subsection{Hippocampal phase precession}

\begin{figure}
\begin{tabular}{lclc}
\textbf{a} && \textbf{b}\\
&
\includegraphics[height=0.12\textheight]{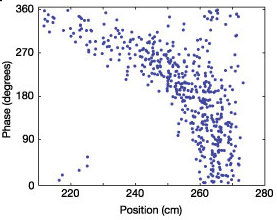}
&&
\includegraphics[height=0.12\textheight]{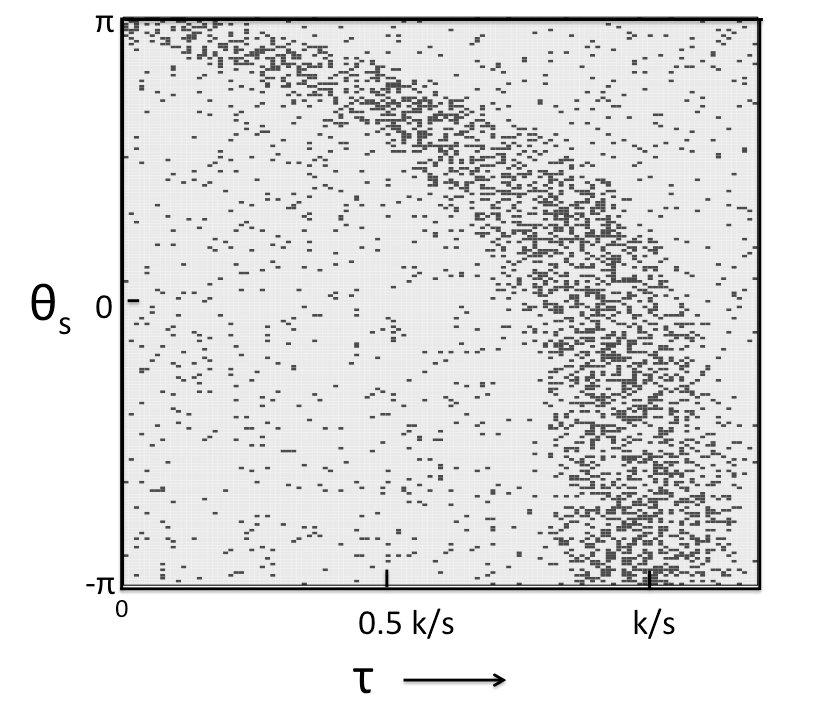}
\end{tabular}
\caption{ 
\textbf{a.}  Neurophysiological data showing phase precession.  Each spike
fired by a place cell is
shown as a function of its position along a linear track (x-axis) and the
phase of local theta (y-axis).  After Mehta, et al., 2002.
		\textbf{b.}
Simulated spikes from a node in the $\T{}$ layer described by
eq.~\ref{eq:bigT} as a function of $\tau$ and local phase $\theta_s$.
The curvature  is a consequence of eq.~\ref{eq:thetas}.
See text for details.
}   \label{fig:ph-pr}
\end{figure}
\nocite{MehtEtal02}

It is clear from eq.~\ref{eq:bigT} that the activity of nodes in the $\T{}$
layer depends on both $\theta_s$ and $\tau$.  Figure~\ref{fig:ph-pr} shows
phase precession data from a representative cell (Fig.~\ref{fig:ph-pr}a,
\cite{MehtEtal02}) and spikes generated from eq.~\ref{eq:bigT}
(Fig.~\ref{fig:ph-pr}b).  The model generates a characteristic curvature for
phase precession, a consequence of  the exponential form of the function
$\Phi$ (eq.~\ref{eq:thetas}). The example cell chosen  in fig.~\ref{fig:ph-pr}
shows roughly the same form of curvature as that generated by the model.
While it should be noted that there is some variability across cells, careful
analyses have led computational neuroscientists to conclude that the canonical
form of phase precession resembles this representative cell.  For instance, a
detailed study of hundreds of phase-precessing neurons \cite{YamaEtal02}
constructed averaged phase-precession plots using a variety of
methods and found a distinct curvature that qualitatively resembles
this neuron.
Because of the analogy between
time and one-dimensional position (eq.~\ref{eq:timespace}), the model yields
the same pattern of  phase precession for time cells and place cells.

\begin{figure}
\includegraphics[width=0.9\columnwidth]{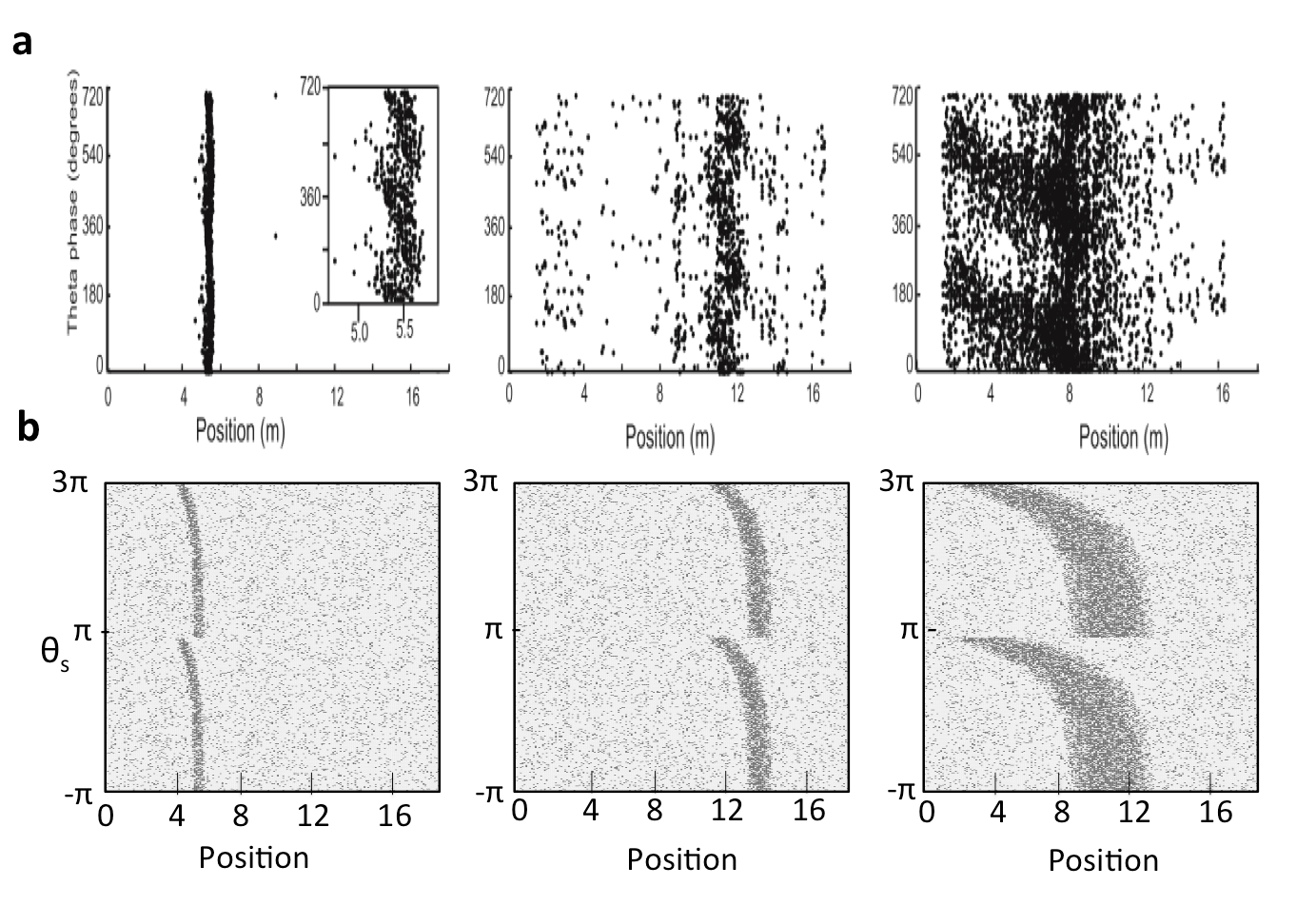}
\caption{ Place cells along the dorsoventral axis of the hippocampus have
place fields that increase in size. \textbf{a.}  The three
panels show the activity of place cells recorded at the dorsal, intermediate
and ventral segments of the hippocampus, when a rat runs along an 18~m track.
After Kjelstrup, et al., (2008). Each spike the cell fired is shown as a
function of position and the \emph{local} theta phase at the cell's location
when it fires (recall that theta phase is not constant across the dorsoventral
axis). Regardless of the width of the place field, neurons at all locations
along the dorsoventral axis phase precess through the same range of {local}
theta phases.  \textbf{b.}
			According to the model, phase precession extends over the same
					range of values of local theta $\theta_s$ regardless of
					the value of $s$, which sets the scale for a particular
					node.  As a consequence, cells with different values of
					$s$ show time/place fields of different size but phase
					precess over the same range of local theta.  For the three
					figures, $s$ values of the nodes are set to $.1$, $.22$,
					and $.7$ respectively, and they are assumed to respond to
							landmarks at location $4$, $11$, and $3$ meters
							respectively from one end of the track.	
\label{fig:Kjel}
	}
\end{figure}

The $\T{}$ layer activity represented in fig.~\ref{fig:ph-pr}a is scale-invariant; note that the $x$-axis is expressed in units of the scale of the node ($k/s$).  
It is known that the spatial scale of place fields changes systematically
along the dorsoventral axis of the hippocampus.  Place cells in the dorsal
hippocampus have place fields of the order of a few centimeters whereas place
cells at the ventral end  have place fields as large as a few meters
(fig.~\ref{fig:Kjel}a) \cite{JungEtal94,KjelEtal08}. 
However, all of them  show the same pattern of precession with respect to
their local theta phase---the phase measured at the same electrode that
records a given place cell (fig.~\ref{fig:Kjel}).  Recall that at any given
moment, the local phase of theta oscillation depends on the position along the
dorsoventral axis \cite{LubeSiap09,PateEtal12}, denoted as the $s$-axis in the
model. 

Figure~\ref{fig:Kjel}a shows the activity of three different place cells  in
an experiment where  rats ran down a long track that extended through open
doors connecting three testing rooms \cite{KjelEtal08}.  The landmarks
controlling a  particular place cell's firing may have been at a variety of
locations along the track. Accordingly, fig.~\ref{fig:Kjel}b shows the
activity of cells generated from the model with different values of $s$ and
with landmarks at various locations along the track (described in the
caption).    
From fig.~\ref{fig:Kjel} it can be qualitatively noted that phase precession
of different cells only depends on the local theta phase and is unaffected by
the spatial scale of firing. This observation is perfectly consistent with the
model.   

\subsection{Prediction of distant rewards \emph{via} phase precession in the
		ventral striatum}

\begin{figure}
	\begin{tabular}{lclc}
		\textbf{a} && \textbf{b}\\
				& \includegraphics[width=0.45\columnwidth]{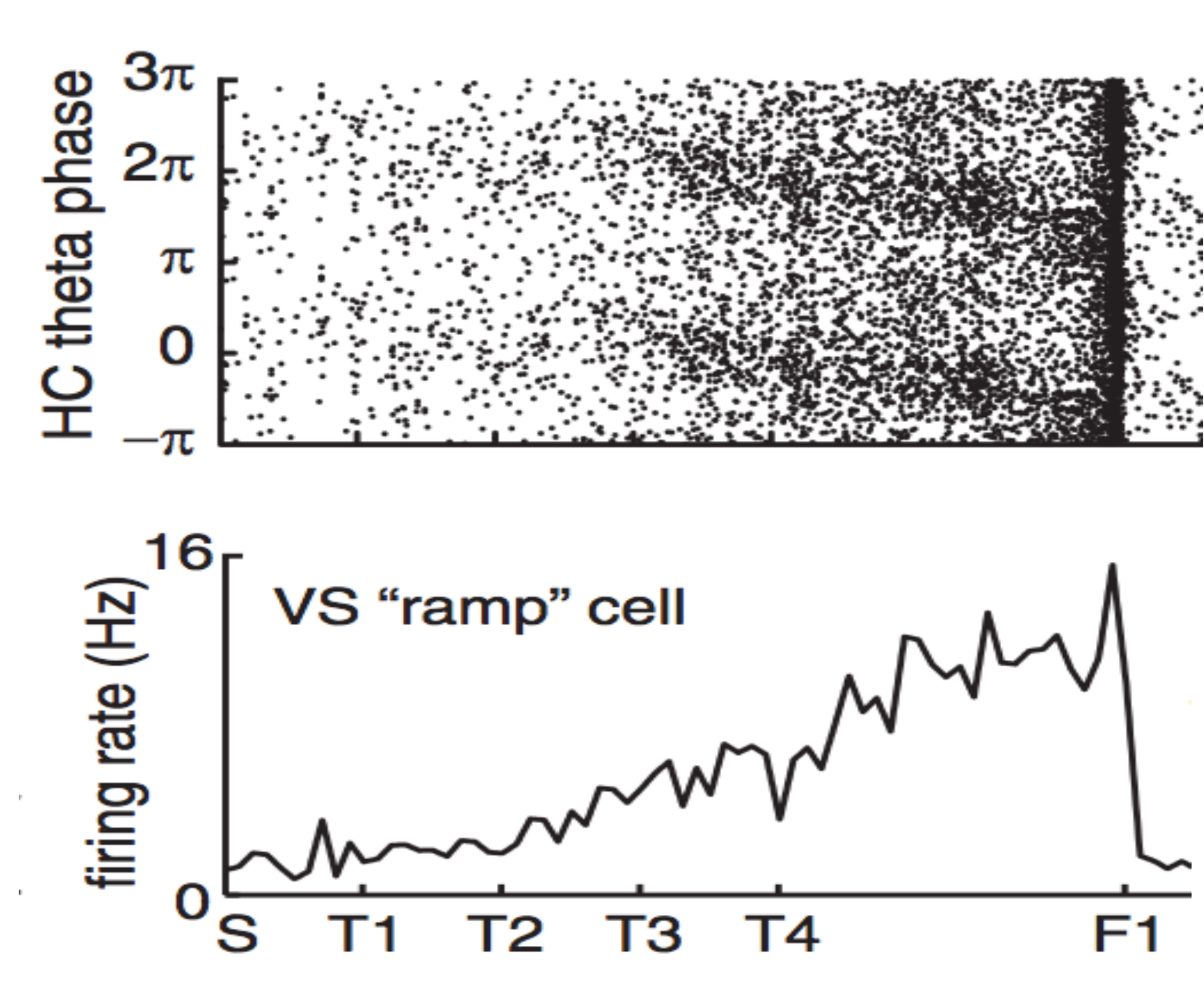}
				&& \includegraphics[width=0.45\columnwidth]{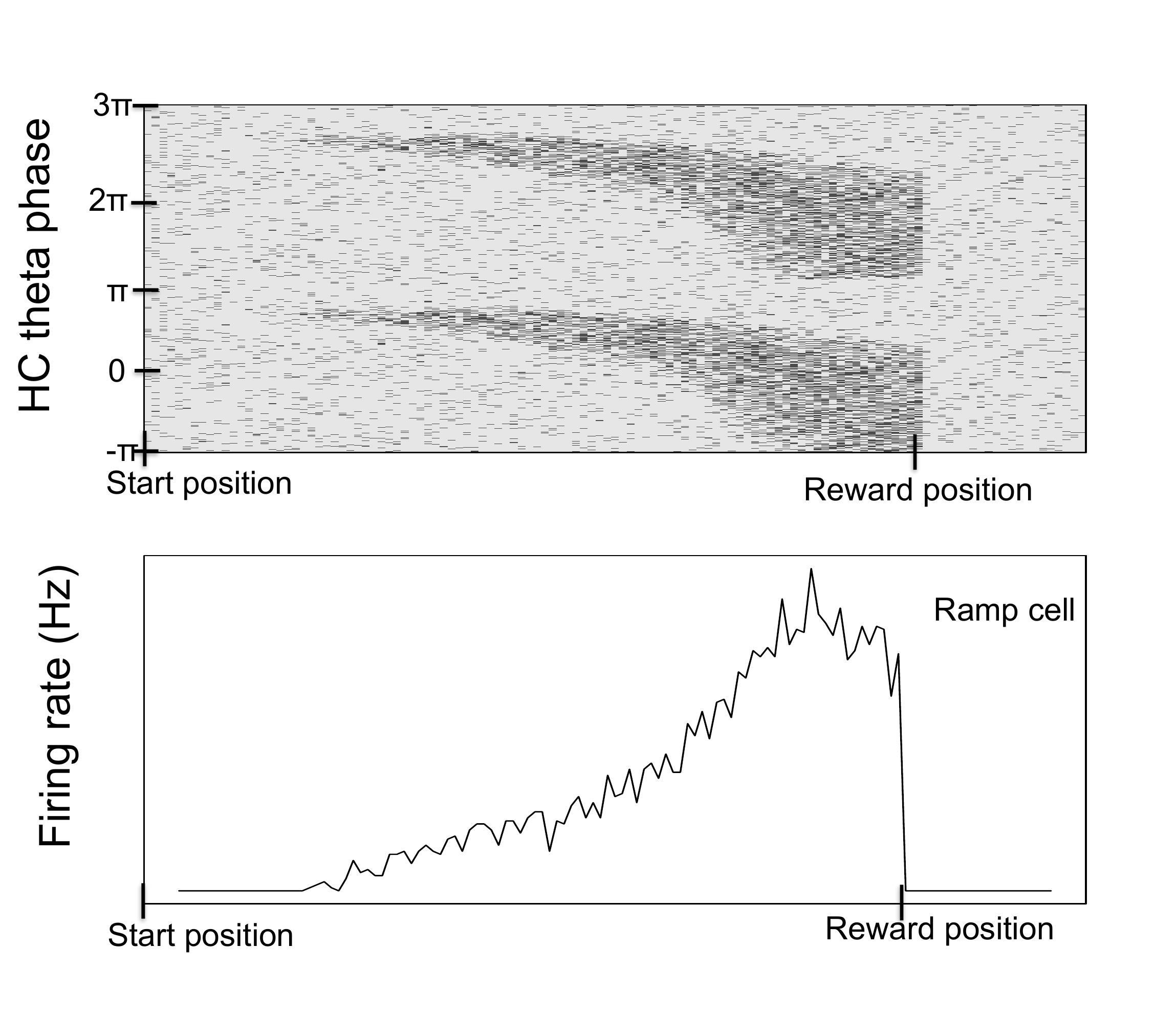}
	\end{tabular}
	\caption{
		\textbf{a.}  A representative ramping cell in the ventral striatum.
				On each trial the animal started the maze at S, made a series
				of turns (T1, T2, etc) and received reward at F1 on 75 percent
				of trials. The total distance between S and F1 is on the order
				of a few meters.  Position along the track is represented
				linearly on the x-axis for convenience.  In the top panel, the
				spikes are shown as a function of theta phase at the dorsal
				hippocampus and position.  The bottom panel shows the firing
				rate as a function of position, which is seen to  gradually
				ramp up towards the reward location.  \textbf{b.}  The
				activity of prediction node generated by the model is plotted
				w.r.t. the reference phase $\theta_o$ and position in the top
				panel, and the the average activity within a theta cycle is
				plotted against position in the bottom panel.  
			\label{fig:MeerRedi}
	}
\end{figure}

We compare the future predictions generated by the model (eq.~\ref{eq:p}) to
an experiment that recorded simultaneously from the hippocampus and nucleus
accumbens, a reward-related structure within the ventral striatum
\cite{MeerEtal11}. Here the rat's task was to learn to make several turns in
sequence on a maze to reach two locations where reward was available.
Striatal neurons fired over long stretches of the maze, gradually ramping up
their firing as a function of distance along the path and terminating at the
reward locations (bottom fig.~\ref{fig:MeerRedi}a). Many striatal neurons
showed robust phase precession relative to the theta phase at the dorsal
hippocampus (top fig.~\ref{fig:MeerRedi}a).  Remarkably, the phase of
oscillation in the hippocampus controlled firing in the ventral striatum to a
greater extent than the phase recorded from within the ventral striatum.   
On trials where there was not a reward at the expected location (F1), there was
another ramp up to the secondary reward location (F2), accompanied again by
phase precession (not shown in fig.~\ref{fig:MeerRedi}a).

This experiment corresponds reasonably well to the conditions assumed in the
derivation of eq.~\ref{eq:p}.  In this analogy, the start of the trial (start
location S) plays the role of the \textsc{cs} and the reward
plays the role of the predicted
stimulus. However, there is a discrepancy between the methods and the
assumptions of the derivation.  The ramping cell (fig.~\ref{fig:MeerRedi}a)
abruptly terminates after the reward is consumed, whereas
		eq.~\ref{eq:p} would gradually decay back towards zero. This is
		because of the way the experiment was set up--there were never two
		rewards presented consecutively.  As a consequence, having just
		received a reward strongly predicts that there will not be a reward in
		the next few moments. In light of this consideration, we force the
		prediction generated in eq.~\ref{eq:p} to be zero beyond the reward
		location and let the firing be purely stochastic.
The top panel of fig.~\ref{fig:MeerRedi}b  shows the spikes generated by model
prediction cells with respect to the reference theta phase $\theta_o$, and the
bottom panel shows the ramping activity computed as the average firing
activity within a complete theta cycle around any moment.

The model correctly captures the qualitative pattern observed in the data.
According to the model, the reward starts being predicted at the beginning of the
track.  Initially, the reward is far in the future, corresponding to a large
value of $\delta$.  As the animal approaches the location of the reward, the
reward moves closer to the present along the $\delta$ axis, reaching zero near
the reward location.   The ramping activity is a consequence of  the
exponential mapping  between $\delta$ and $\theta_o$ in eq.~\ref{eq:thetas}.
Since the proportion of the theta cycle devoted to large values of $\delta$ is
small, the firing rate averaged across all phases will be small, leading to an
increase in activity closer to the reward.

\subsection{Testable properties of the mathematical model}

\begin{figure}
\includegraphics[width=0.8\columnwidth]{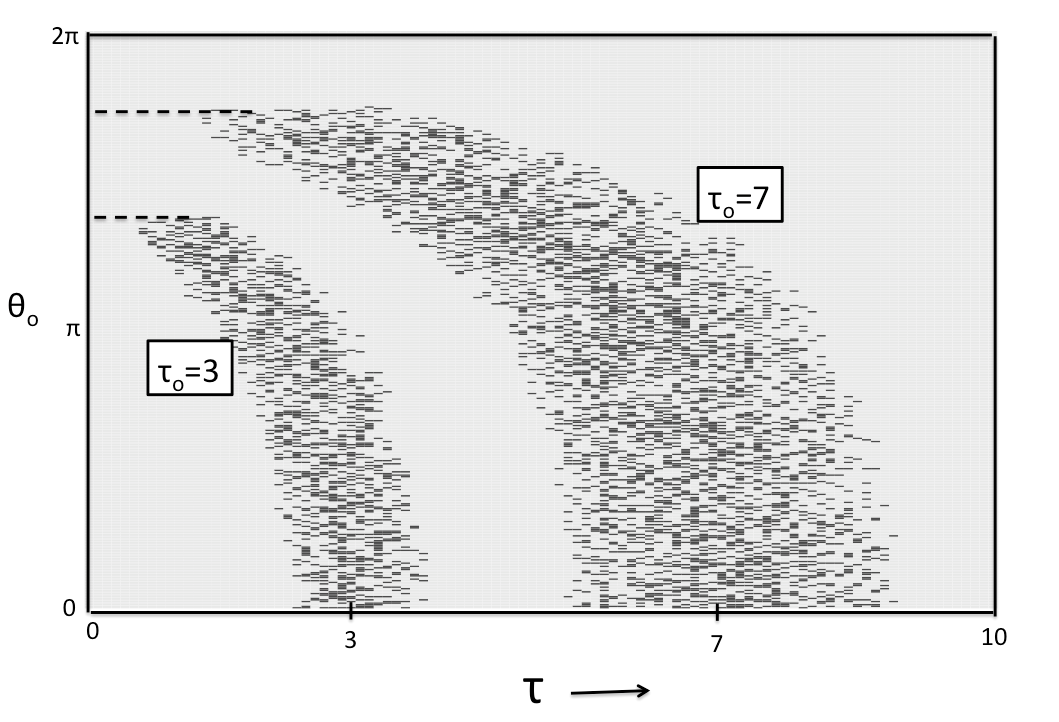}
\caption{Changing $\tau_o$ affects the phase at which prediction cells  start
		firing.  At early times,  the magnitude of translation required to
		predict the $\tau_o=3$ outcome  is smaller than that required to
		predict the $\tau_o=7$ outcome. Consequently, the cell begins to fire
		at a larger $\theta_o$  for $\tau_o=7$. Parameter values are the same
		as the other figures as given in the beginning of this section, except
		for clarity the background probability of spiking has been set to
		zero.  \label{fig:phaseentry}}
\end{figure}

Although the model aligns reasonably well with known properties of theta phase
precession, there are a number of features of the model  that have, to our
knowledge, not yet been evaluated.  At a coarse level, the correspondence
between time and one-dimensional space implies that time cells should exhibit
phase precession with the same properties as place cells. While phase
precession has been extensively observed and characterized in hippocampal
place cells, there is much less evidence for phase precession in hippocampal
time cells (but see
\cite{PastEtal08}).

According to the model, the pattern of phase precession is related to the
distribution of $s$ values represented along the dorsoventral axis. While  it
is known that a range of spatial scales are observed along the dorsoventral
axis, their actual distribution is not known.  The Weber-Fechner scale of
eq.~\ref{eq:thetas} is a strong prediction of the framework developed here.
Moreover, since $\Phimax/\Phio = s_o/s_N$, the ratio of the largest to
smallest scales represented in the hippocampus places constraints on the form
of phase precession. The larger this ratio, the larger will be the value of
$b$ in eq.~\ref{eq:thetas}, and the curvature in the phase precession plots
(as in fig.~\ref{fig:ph-pr}) will only emerge at larger values of the local
phase $\theta_s$.  Neurophysiological observation of this ratio could help
evaluate the model. 

The form of $\p{\delta}$ (eq.~\ref{eq:p}) leads to several distinctive
features in the pattern of phase precession of the nodes representing future
prediction.  It should be possible to observe phase precession for cells that
are predicting any stimulus, not just a reward. In addition, the model's
assumption that a timeline of future predictions is aligned with global theta
phase 
 has interesting measurable consequences.  Let's reconsider the thought
experiment from the previous section (fig.~\ref{fig:futurepred}), where a
stimulus predicts an outcome after a delay $\tau_o$. Immediately after the
stimulus is presented, the value of $\delta$  at which the prediction peaks is
monotonically related to $\tau_o$. Since $\delta$ is monotonically related to
the reference phase $\theta_o$, the prediction cells will begin to fire at later
phases  when $\tau_o$ is large, and as time passes, they will fire at earlier
and earlier phases all the way untill $\theta_o=0$.  In other other words, the
entry-phase (at which the firing activity begins) should depend on $\tau_o$, the
prediction timescale. This is illustrated in fig.~\ref{fig:phaseentry} with
$\tau_o=3$ and $\tau_o=7$, superimposed on the same graph to make visual
comparison easy. The magnitude of the peak activity would in general depend on
the value of $\tau_o$ except when $w=2$ (as assumed here for visual clarity).
Experimentally manipulating the reward times  and studying the phase
precession of prediction cells could help test this feature.

\section{Discussion}
\label{sec:discussion}

This paper presented a neural hypothesis for implementing translations of
temporal and 1-d spatial memory states so that future events can be quickly
anticipated without destroying the current state of memory.  The hypothesis
assumes that time cells and place cells observed in the hippocampus represent
time or position as a result of a two-layer architecture that encodes and
inverts the Laplace transform of external input. It also assumes that
sequential translations to progressively more distant points in the future
occur within each cycle of theta oscillations.  Neurophysiological
constraints were imposed as phenomenological rules rather than as emerging
from a detailed circuit model.  Further, imposing scale-invariance and
coherence in translation across memory nodes resulted in Weber-Fechner
spacing for the representation of both the past (spacing of $s_n$ in the
memory nodes) and the future (the relationship between $\delta$ and
$\theta_o$).  Apart from providing cognitive flexibility in accessing a
timeline of future predictions at any moment, the computational mechanism
described qualitative features of phase precession in the hippocampus and in
the ventral striatum.  Additionally, we have also pointed out certain
distinctive features of the model that can be tested with existing technology.

\subsection{Computational Advantages}
The  property of the $\T{}$ layer that different nodes represent the stimulus
values from various delays (past moments) is reminiscent of a shift register
(or delay-line or synfire chain). However, the two layer network encoding and
inverting the Laplace transform of stimulus has several significant
computational advantages over a shift register representation. 

(i) In the current two-layer network, the spacing of $s$ values of the nodes
can be chosen freely. By choosing exponentially spaced $s$-values
(Weber-Fechner scaling) as in eq.~\ref{eq:thetas}, the $\T{}$ layer can
represent memory from exponentially long timescales compared to a shift
register with equal number of nodes, thus making it extremely
resource-conserving. Although information from longer timescales is more
coarse-grained, it turns out that this  coarse-graining is optimal to
represent and predict long-range correlated signals \cite{ShanHowa13}. 

(ii) The memory representation of this two layer network is naturally
scale-invariant (eq.~\ref{eq:old_big_T}). To construct a scale-invariant
representation from a shift register, the shift register would have to be
convolved with a scale-invariant coarse-graining function at each moment,
which would be computationally very expensive. Moreover, it turns out that any
network that can represent such scale-invariant memory  can be identified with
linear combinations of multiple such two layer networks \cite{Shan15}. 

(iii)
Because
translation can be trivially performed when we have access to the Laplace
domain, the two layer network enables translations by an amount
$\delta$ without sequentially visiting the intermediate states $<\delta$. This can be
done by directly changing the connection strengths locally
between the two layers as prescribed by diagonal $\Rd$ operator for any chosen
$\delta$.\footnote{In this paper we considered sequential translations of
various values of $\delta$, since the aim was to construct an
entire future timeline rather than to discontinuously jump to a
distant future state.} Consequently the physical time taken for the
translation can be decoupled from the magnitude of translation. One could
imagine a shift register performing a translation operation by an amount
$\delta$ either by shifting the values  sequentially from one node to the next
for $\delta$ time steps or by establishing non-local connections between far
away nodes. The latter would make the computation very cumbersome because it
would require every node in the register to be connected to every other node
(since this should work for any $\delta$), which is in stark contrast with the
local connectivity required by our two layer network to perform any
translation. 

Many previous neurobiological models of  phase precession have been proposed
\cite{LismJens13,MehtEtal02,BurgEtal07,Hass12}, and many assume that
sequentially activated place cells firing within a theta cycle result from
direct connections between those cells \cite{JensLism96}, not unlike a synfire
chain. Although taking advantage of the Laplace domain in the two
layer network to perform  translations is not the only possibility, it seems
to be computationally powerful compared to the obvious alternatives.

\subsection{Translations without theta oscillations}

Although this paper focused on sequential translation within a theta cycle,
translation may also be accomplished \emph{via} other neurophysiological
mechanisms. Sharp wave ripple (SRW) events last for about 100~ms and are
often accompanied by replay events--sequential firing of place cells
corresponding to locations different from the animal's current
location \cite{DaviEtal09,DragTone11,FostWils06,PfeiFost13,JadhEtal12}.  Notably,
experimentalists have also observed preplay events during SWRs, sequential activation of
place cells that correspond to trajectories that have never been
previously traversed, as though the animal is planning  a future path
\cite{DragTone11,OlafEtal15}.
Because untraversed trajectories could not have been used to learn and build
sequential associations between the place cells along the trajectory, the
preplay activity could potentially be a result of a translation operation on
the overall spatial memory representation.  

Sometimes during navigation, a place cell corresponding to a distant goal
location gets activated \cite{PfeiFost13}, as though a finite distance
translation of the memory state has occurred.  More interestingly, sometimes a
reverse-replay is observed in which place cells are activated in reverse order
spreading back from the present location \cite{FostWils06}. This is suggestive
of translation into the past (as if $\delta$ was negative), to implement a
memory search.  In parallel, there is behavioral evidence from humans that
under some circumstances memory retrieval consists of a backward scan through
a temporal memory representation \cite{Hack80,Hock84,SingEtal15} (although
this is not neurally linked with SWRs).  Mathematically, as long as the
appropriate connection strength changes prescribed by the $\Rd$ operator can
be specified, there is no reason translations with negative $\delta$ or
discontinuous shift in $\delta$ could not be accomplished in this framework.
Whether these computational mechanisms are reasonable in light of the
neurophysiology of sharp wave ripples is an open question.

\subsection{Multi-dimensional translation}

This paper focused on  translations along one dimension. However it would be
useful to extend the formalism to multi-dimensional translations. When a rat
maneuvers through an open field rather than a linear track, phase precessing
2-d place cells are observed \cite{SkagEtal96}.  Consider the case of an
animal approaching a junction along a maze where it has to either turn left or
right. Phase precessing cells in the hippocampus indeed predict the direction
the animal will choose in the future \cite{JohnRedi07}.  In order to
generalize the formalism to 2-d translation, the nodes in the network model
must not be indexed only by $s$, which codes their distance from a landmark,
but also by the 2-d orientation along which distance is calculated.  The
translation operation must then specify not just the distance, but also
the instantaneous direction as a function of the theta phase.  Moreover,
if translations could be performed on multiple non-overlapping
trajectories simultaneously, multiple paths could be searched in
parallel, which  would be very useful for efficient decision making.

\subsection{Neural representation of predictions}

The computational function of $\p{\delta}$ (eq.~\ref{eq:p}) is to represent an
ordered set of events predicted to occur in the  future.  Although we focused
on ventral striatum here because of the availability of phase precession data
from that structure, it is probable that many brain regions represent future
events as part of a circuit involving frontal cortex and basal ganglia, as
well as the hippocampus and striatum
\cite{SchuEtal97,FerbShap03,TanaEtal04,FeieEtal06,MainKepe09,TakaRoes11,YounShap11a}.
There is evidence that theta-like oscillations coordinates the activity in
many of these brain regions
\cite{JoneWils05,LansEtal09,vanWEtal10,FujiBuzs11}.    For instance, 4~Hz
oscillations show phase coherence between the hippocampus, prefrontal cortex
and ventral tegmental area (VTA), a region that signals the presence of
unexpected rewards \cite{FujiBuzs11}.  A great deal of experimental work has
focused on the brain's response to future rewards, and indeed the
phase-precessing cells in fig.~\ref{fig:MeerRedi} appear to be predicting the
location of the future reward.  The model suggests that $\p{\delta}$ should
predict any future event, not just a reward. Indeed, neurons that appear to
code for predicted stimuli have been observed in the primate inferotemporal
cortex \cite{SakaMiya91} and prefrontal cortex \cite{RainEtal99}.  Moreover,
theta phase coherence between prefrontal cortex and hippocampus are
essential for learning the temporal relationships between stimuli
\cite{BrinMill15}. 
So, future predictions could be widely distributed throughout the brain.


%

\end{document}